\documentclass[a4paper,fleqn]{cas-sc}

\usepackage{chngcntr}
\usepackage{booktabs}
\usepackage{ulem}
\usepackage[numbers]{natbib}
\usepackage[utf8]{inputenc}
\usepackage[T1]{fontenc}
\newtheorem{thm}{Theorem}
\newtheorem{ex}{Example}

\usepackage[utf8]{inputenc}
\usepackage{amsmath,amssymb}
\usepackage{float}            
\usepackage{algorithm}        
\usepackage{algpseudocode}    

\def\tsc#1{\csdef{#1}{\textsc{\lowercase{#1}}\xspace}}
\tsc{WGM}
\tsc{QE}
\tsc{EP}
\tsc{PMS}
\tsc{BEC}
\tsc{DE}

\begin{document}
\let\WriteBookmarks\relax
\def\floatpagepagefraction{1}
\def\textpagefraction{.001}
\shorttitle{Boolean-network simplification and rule fitting to unravel chemotherapy resistance in non-small cell lung cancer}
\shortauthors{Espinoza et al. 2025}

\title [mode = title]{Boolean-network simplification and rule fitting to unravel chemotherapy resistance in non-small cell lung cancer}                    

\author[1]{Alonso Espinoza}[type=editor,
                        auid=000,bioid=1,
                        orcid=0009-0009-4643-5003]
\cormark[1]
\ead{alonespinoza@alumnos.uai.cl}

\credit{Conceptualization of this study, Methodology, Software}

\author[1,2]{Eric Goles}
\ead{eric.chacc@uai.cl}

\credit{Data curation, Writing - Original draft preparation}

\affiliation[1]{organization={Facultad de Ingenieria y Ciencias, Univ. Adolfo Ibañez},
                addressline={Av. Diagonal Las Torres 2700}, 
                postcode={7910000}, 
                postcodesep={}, 
                city={Santiago},
                country={Chile}}

\affiliation[2]{organization={Millennium Nucleus for Social Data Science (SODAS)},
                city={Santiago},
                country={Chile}}

\author[1]{Marco Montalva-Medel}
\ead{marco.montalva@uai.cl}

\cortext[cor1]{Corresponding author}

\begin{abstract}
Boolean networks are powerful frameworks for capturing the logic of gene-regulatory circuits, yet their combinatorial explosion hampers exhaustive analyses. Here, we present a systematic reduction of a 31-node Boolean model that describes cisplatin- and pemetrexed-resistance in non-small-cell lung cancer to a compact 9-node core that exactly reproduces the original attractor landscape. The streamlined network shrinks the state space by four orders of magnitude, enabling rapid exploration of critical control points, rules fitting  and candidate therapeutic targets. Extensive synchronous and asynchronous simulations confirm that the three clinically relevant steady states and their basins of attraction are conserved and reflect resistance frequencies close to those reported in clinical studies. The reduced model provides an accessible scaffold for future mechanistic and drug-discovery studies.
\end{abstract}



\begin{keywords}
Drug resistance \sep NSCLC \sep Boolean network \sep Steady states \sep Limit cycles \sep Update schedule
\end{keywords}

\maketitle

\section{Introduction}

Non-small cell lung cancer (NSCLC) is a type of cancer that accounts for approximately 85\% of all lung cancer cases \cite{gridelli2015non}. It is a heterogeneous disease that includes several subtypes such as adenocarcinoma, squamous cell carcinoma, and large cell carcinoma. The primary risk factor for developing NSCLC is smoking, although other factors such as radon exposure and air pollution have also been identified \citep{gridelli2015non}. Most patients are diagnosed at advanced stages due to the lack of effective screening programs and the late onset of clinical symptoms, resulting in a generally unfavorable prognosis.

This study focuses on the dynamical analysis of a Boolean genetic network designed to investigate drug resistance mechanisms in NSCLC. This network, originally proposed in a recent research, includes interactions between long non-coding RNAs, microRNAs, and various transcription factors that collectively influence resistance to chemotherapeutic treatments such as cisplatin and pemetrexed \citep{gupta2024dynamic}. The structure and dynamics of the network were modeled using a Boolean network approach, which allows the simulation of genetic regulation through binary states (active/inactive), thus facilitating the study of network stability and response under different update schemes.

Boolean networks are widely recognized for their ability to capture gene interactions and the overall dynamic behavior of genetic regulatory networks, making them a popular model for inferring model structure from gene expression data. These approaches are particularly useful in situations where gene expression measurements are noisy and can lead to inconsistent observations \citep{lahdesmaki2003learning}.

Boolean genetic networks, which simplify the complexity of biological interactions into binary relationships, emerge as powerful tools in decoding the multifaceted dynamics of drug resistance. By modeling these interactions within the context of NSCLC, this study aims not only to identify key network configurations that promote resistance but also to evaluate how modifications in these configurations can alter cellular phenotypes. This approach can reveal novel therapeutic targets and offer insights into overcoming chemotherapy resistance.

Analyzing all deterministic dynamics of a Boolean regulatory network is a challenging problem, as it grows exponentially with the number of nodes. In this context, mathematical and computational tools, such as the notion of alliance and equivalence classes of update schemes, are introduced to study dynamic robustness under different update schemes \citep{aracena2011combinatorics,goles2013deconstruction}. These tools are essential for discovering new information related to the model's attractors, considering all possible deterministic dynamics (i.e., not only the synchronous dynamics that is typically used in these models).

Moreover, studies such as the one conducted by Aracena et al. \citep{aracena2013number} have demonstrated the efficacy of update scheme tools in revealing the complex dynamics of Boolean networks. Although these tools are not specifically designed to evaluate stability in the traditional sense of stability analysis, they offer valuable resources for investigating how the sequence of updates influences the overall stability of systems modeled as Boolean networks. This is particularly useful in systems whose dynamics are sensitive to the order in which their components are updated.

Therefore, it is proposed that the application and comparison of different update strategies in the Boolean network model will enhance the identification of minimal network configurations essential for obtaining cellular phenotypes in non-small cell lung cancer. To achieve this, it is proposed to reconstruct the Boolean network model in logical terms, simulate the model synchronously to confirm the reported steady states (or fixed points), deconstruct the Boolean network model, comparatively evaluate update strategies for Boolean network models, and identify minimal network configurations necessary for reproducing steady states.

\section{Methods}
\subsection{Data Acquisition}
The Boolean network used in this study originates from the scientific article "A dynamic Boolean network reveals that the BMI1 and MALAT1 axis is associated with drug resistance by limiting miR-145-5p in non-small cell lung cancer" \citep{gupta2024dynamic}.

\subsection{Data Analysis}
The R programming language (version 4.3.3) \cite{rsoftware} was used to work with the data, utilizing the R package BoolNet, which is a tool designed for the simulation and analysis of genetic regulatory networks and other biological systems modeled using Boolean networks \citep{mussel2010boolnet}.

The Python programming language (version 3.10.5) \cite{python} was used to generate simulation programs for the different update schemes and to obtain the basins of attraction for each of them. These were also used to simulate the regulation fitting algorithm. Also used to simulate the regulation matching algorithm, these algorithms are available on GitHub 
(\href{https://github.com/aer-neo/Different-asynchronous-update-scheme-tools}
{https://github.com/aer-neo/Different-asynchronous-update-scheme-tools}) 
and 
(\href{https://github.com/aer-neo/Fitting-Boolean-Rules-in-GNR}
{https://github.com/aer--neo/Fitting-Boolean-Rules-in-GNR}).

\subsection{Boolean networks}
Boolean Networks (BNs) are simple yet powerful models introduced decades ago \citep{kauffman1969metabolic} to represent complex systems. In these models, biological components are represented by binary states ("0" or "1"), which helps to simplify dynamic processes and offer insights into their behavior. BNs have been used extensively to study Gene Regulatory Networks (GRNs) across various species (e.g., \citep{albert2003topology,goles2013deconstruction,ruz2014dynamical,montalva2021lac,vivanco2024dynamical}). A formal BN consists of a finite directed graph where nodes correspond to biological components, each having a discrete state of either 0 (inactive) or 1 (active), and the edges represent the interactions between these components. The states of the nodes are updated according to Boolean functions, following a predefined update schedule. For a network with $n$ nodes, the update function can be defined as $s : \{1, ..., n\} \to \{1, ..., n\}$, where $s(i) < s(j)$ indicates that node $i$ is updated before node $j$. These update schemes have proven to be useful for analyzing various biological applications (e.g., \citep{albert2003topology,goles2013deconstruction,ruz2014dynamical,montalva2021lac,vivanco2024dynamical}) as well as for developing new theoretical results in discrete mathematics (e.g., \citep{ruz2013preservation,goles2018block,perrot2020maximum}). In this work, we will use the following ones:
\begin{itemize}
 \item (Deterministic) Synchronous update schedule or parallel update: $s(k)=1$, for all node $k$, at each time step. In other words, at each time step, all nodes are updated at the same time (simultaneously). Notation: $s=(1,2,...,n)$.
 \item (Deterministic) Asynchronous update schedule or block sequential update schedule: if $A_1$,...,$A_k$ is a partition of the set of nodes $\{1,...,n\}$, $1< k\leq n$, then $s(j)=i$, for all $j\in A_i$, at each time step. In other words, at each time step, the blocks of nodes $A_i$ are updated one by one (or sequentially) and the nodes inside them in a parallel way. Notation: $s=(A_1)(A_2)\cdots (A_k)$.
\end{itemize}

The above implies that a specific dynamic behavior of the network is determined by the choice of an update schedule. This is typically represented by a state transition graph, where each node corresponds to a configuration (or vector of states) of the network’s binary states, and the arcs represent transitions between configurations at each time step, resulting from the application of local Boolean functions according to the update schedule considered. Additionally, since there are $2^n$ possible configurations, the network will always evolve toward a subset of configurations known as attractors. These attractors can be either fixed points (or steady states), where the configuration remains unchanged after applying the Boolean functions, or limit cycles with a period $l > 1$, consisting of $l$ configurations $\{x(0), x(1), ..., x(l-1)\}$, where each configuration $x(t)$ transitions to $x(t+1(\mod l))$. The attraction basin of an attractor refers to the set of configurations that eventually lead to that attractor.

In this context, a 31-node Boolean Network model for the study of drug resistance mechanisms in NSCLC was introduced in \citep{gupta2024dynamic}, where the authors employed a synchronous update scheme and that we reconstruct in this manuscript with some small adjustments (see Figure \ref{FIG:A2}, the corresponding local Boolean functions in Table \ref{Atbl2} and the reconstruction details in section 3.1 in results) and then we reduce it (details in section 4.1 in discussion) along with considering all the deterministic update schedules explained at the beginning of this section. This represents a large family of schemes that grow exponentially as the network size increases \citep{ruz2013preservation}. Specifically, the number $T_n$ of deterministic update schedules corresponding to a directed graph with $n$ vertices is given by:
\begin{equation*}\label{Tn}
T_n=\sum_{k=0}^{n-1} \binom{n}{k}T_k ,
\end{equation*}
where $T_0\equiv 1$. Additionally, if we consider that the dynamics obtained by each of these $T_n$ schemes generates $2^n$ possible configurations, the computational effort required for an exhaustive analysis of all these dynamics (which we will conduct in this study) becomes even more demanding. In practice, such an analysis can only be performed for small values of $n$ but, in the following section, we will outline the key concepts of update digraph theory which significantly reduces these computational costs. 

\subsection{Update digraphs and different dynamics in boolean networks}
In this section, we will provide a summary of the key concepts and results from \citep{aracena2009robustness, aracena2011combinatorics, aracena2013number, aracena2013tournament}, which facilitate the classification of deterministic update schemes that produce identical dynamics. Let $G = (V, A)$ be a directed graph, $s$ an update schedule, and the labeling function $lab_s : A \to \{-, +\}$ defined as:
\[
\forall (i,j) \in A, \, lab_s(i,j) = \begin{cases} + & \text{if } s(i) \geq s(j), \\ - & \text{if } s(i) < s(j). \\ \end{cases}
\]
 
By assigning the labels $+$ or $-$ to each arc $(i,j)$ in $A$, we obtain a labeled directed graph $(G, lab)$. When $lab = lab_s$ for some update schedule $s$, the labeled graph $(G, lab_s)$ is referred to as the update digraph. It is important to note that while every update digraph is a labeled graph, the reverse is not necessarily true. In \citep{aracena2011combinatorics}, the following characterization of update digraphs was established:

\begin{thm}\label{teo_charac-ud} 
A labeled digraph is an update digraph if and only if reversing the direction of the arcs labeled $-$ results in a new digraph (possibly a multigraph) that contains no cycles with $-$ labeled arcs. 
\end{thm}

Additionally, in \citep{aracena2009robustness} was proven the following result:

\begin{thm}\label{teo_eqdyn} 
Let $N_1 = (G, F, s_1)$ and $N_2 = (G, F, s_2)$ be two Boolean networks (i.e., they share the same graph $G$ and global function $F$, but differ only in their update schemes $s_1$ and $s_2$). If their update digraphs are identical, i.e., $(G, lab_{s_1}) = (G, lab_{s_2})$, then $N_1$ and $N_2$ exhibit the same dynamics. 
\end{thm}

Using Theorem \ref{teo_eqdyn}, we can define equivalence classes of update schemes that lead to the same update digraph, and therefore produce identical dynamics:

\begin{equation*}\label{def_eq_class}
[s]_{G}=\{s': \ (G,lab_s)=(G,lab_{s'})\}
\end{equation*}

This formulation establishes a bijection between update digraphs and these equivalence classes, which partition the set of all $T_n$ update schemes. Consequently, this allows the study of distinct network dynamics by focusing only on a subset of update schemes, specifically those that serve as representatives of each equivalence class. An algorithm for determining these representative schemes for each class was introduced in \citep{aracena2013number}.

In the following example, we will revisit and elaborate on all of these concepts and results.

\begin{ex}\label{ex_1}
Consider the digraph $G=(V,E)$ where $V=\{A,B,C\}$ and $E=\{(A,B),(B,C),(C,B),(B,A)\}$.
\end{ex}

\begin{figure}[ht]
	\centering
	\includegraphics[width=1\textwidth]{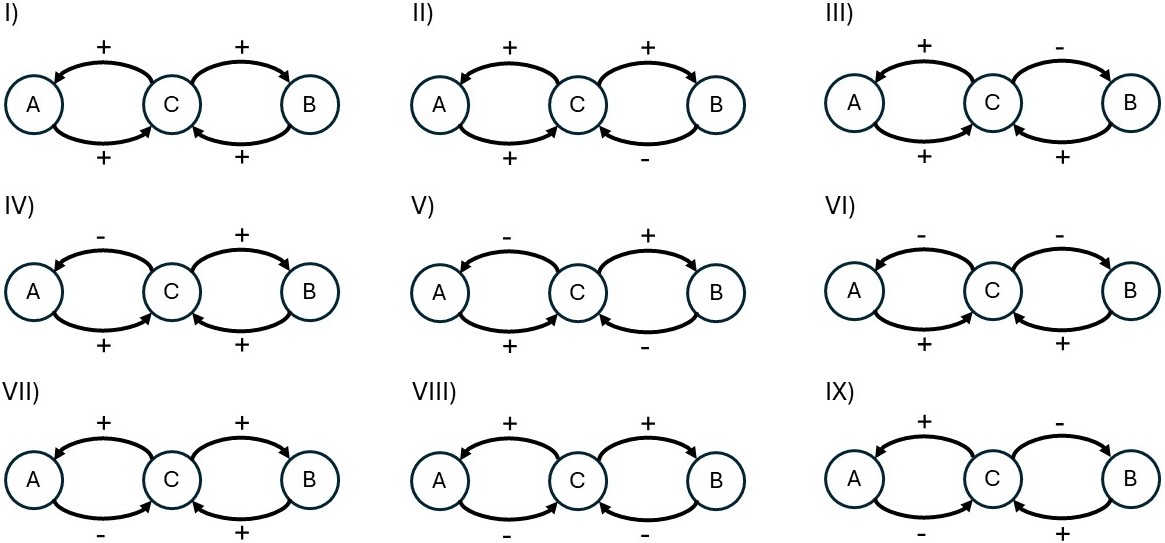}
	\caption{All the update digraphs associated to the digraph $G$ of Example \ref{ex_1} where I) is obtained with $s_1$, II) is obtained with $s_2$, III) is obtained with $s_3$, IV) is obtained with $s_4$, V) is obtained with $s_5$, VI) is obtained with $s_6$ (and also with $s_{10}$ and $s_{13}$), VII) is obtained with $s_7$, VIII) is obtained with $s_8$ (and also with $s_{11}$ and $s_{12}$), and IX) is obtained with $s_9$.}
	\label{FIG:1}
\end{figure}

The following can be easily verified:
\begin{enumerate}
 \item There exist a total of $T_3 = 13$ deterministic update schedules: $s_1=(A,B,C)$, $s_2=(C)(A,B)$, $s_3=(A,B)(C)$, $s_4=(B,C)(A)$, $s_{5}=(C)(B)(A)$, $s_6=(B)(A,C)$, $s_7=(A)(B,C)$, $s_8=(A,C)(B)$, $s_{9}=(A)(B)(C)$, $s_{10}=(B)(C)(A)$, $s_{11}=(A)(C)(B)$, $s_{12}=(C)(A)(B)$ and $s_{13}=(B)(A)(C)$. Notice that $s_1$ is the parallel scheme.
 
 \item There are a total of $2^4=16$ possible labeled digraphs (two possible labels for each arc). From them, only the 9 labeled digraphs shown in Figure \ref{FIG:1} satisfy the Theorem \ref{teo_charac-ud}. This means that some of the update digraphs generated by the 13 previous schemes necessarily repeat themselves and, therefore, the 9 equivalence classes shown in Table \ref{tbl1} are produced.
\end{enumerate}

\begin{table}[width=.9\linewidth,cols=4,pos=ht]
\caption{The different equivalence classes associated to $G$ and the schemes that each one has, where the representative schemes for each class that have been considered are $s_i$, $i=1,...,9$.}
\label{tbl1}
\begin{tabular*}{\tblwidth}{@{} LLLLLLLLL@{} }
\toprule
 $[s_{1}]_F$ & $[s_{2}]_F$ & $[s_{3}]_F$ & $[s_{4}]_F$ & $[s_{5}]_F$ & $[s_{6}]_F$ & $[s_{7}]_F$ & $[s_{8}]_F$ & $[s_{9}]_F$ \\
\midrule
(A,B,C) & (B)(A,C) & (A,C)(B) & (B,C)(A) & (B)(C)(A) & (C)(A,B) & (A)(B,C) & (A,B)(C) & (A)(C)(B) \\
&  &  &  &  & (C)(A)(B) &  & (A)(B)(C) & \\
&  &  &  &  & (C)(B)(A) &  & (B)(A)(C) & \\
\bottomrule
\end{tabular*}
\end{table}

\begin{enumerate}
 \item[3.] Therefore, if one wishes to exhaustively study all the distinct dynamics that a BN with associated digraph $G$ can have, it is not necessary to use all 13 possible update schemes. Instead, only 9 of them are needed (a reduction of nearly 31\%), specifically, $s_i$, $i = 1,...,9$. These 9 must be representative in the sense that they define distinct equivalence classes. Consequently, the BN will have at most 9 distinct dynamics. Figure \ref{FIG:A1} shows a concrete example of all the different dynamics that a BN with digraph $G$ and local boolean functions given by:
 
 \begin{center}
 \begin{tabular}{l}
$x_A =x_C$\\
$x_B =x_C $\\
$x_C =x_A\wedge x_B$
\end{tabular}
\end{center}
can have. Note that having the information of all the different dynamics, one can discuss any dynamic property of the network, for instance, robustness, which could be understood in the sense of whether the observed dynamics retain certain attractors. In this sense, the BN of this example would be considered robust.
\end{enumerate}

In the theory of update digraphs presented in \citep{aracena2009robustness, aracena2011combinatorics, aracena2013number, aracena2013tournament}, additional properties are established that explain why the reduction in the number of schemes required to analyze the distinct dynamics of a Boolean network is typically much more significant than the 31\% reduction observed in this example, particularly as the network size increases.

In this way, we will explore what happens beyond the parallel scheme, which may be biased \citep{harvey1997time} or vulnerable to perturbations in the schemes, as is the case with certain families of Boolean networks, such as elementary cellular automata. In these networks, some can exhibit strong robustness, known as block invariants, while others may not \citep{goles2015block, goles2018block, perrot2020maximum, ruivo2020maximum}.

\section{Results}

\subsection{Network Reconstruction}

The gene regulatory network used in this study was reconstructed from the work of Gupta et al. \cite{gupta2024dynamic}. Originally designed to investigate mechanisms of drug resistance in non-small cell lung cancer (NSCLC), this network includes 31 nodes representing key biological components such as long non-coding RNAs, microRNAs, and transcription factors. These interact to influence resistance to chemotherapies such as cisplatin and pemetrexed. The interactions between nodes were modeled using Boolean functions detailed in Table \ref{Atbl2} of the original article, while the complete network is illustrated in Figure \ref{FIG:A2}.

Following reconstruction, initial simulations under a synchronous update scheme revealed the presence of four terminal phenotypes or stable states: proliferation, drug resistance, senescence, and apoptosis, along with three limit cycles. These cycles correspond to periodic oscillations in node configurations and are described in Table \ref{Atbl3}. The original article from which the network was derived does not indicate whether these limit cycles fulfill a biological function; therefore, they are assumed to be spurious cycles. The initial analysis validated that the reconstructed Boolean functions preserve the original state configurations, confirming the accuracy of the reconstructed network.

\begin{figure}[ht]
	\centering
	\includegraphics[width=0.9\textwidth]{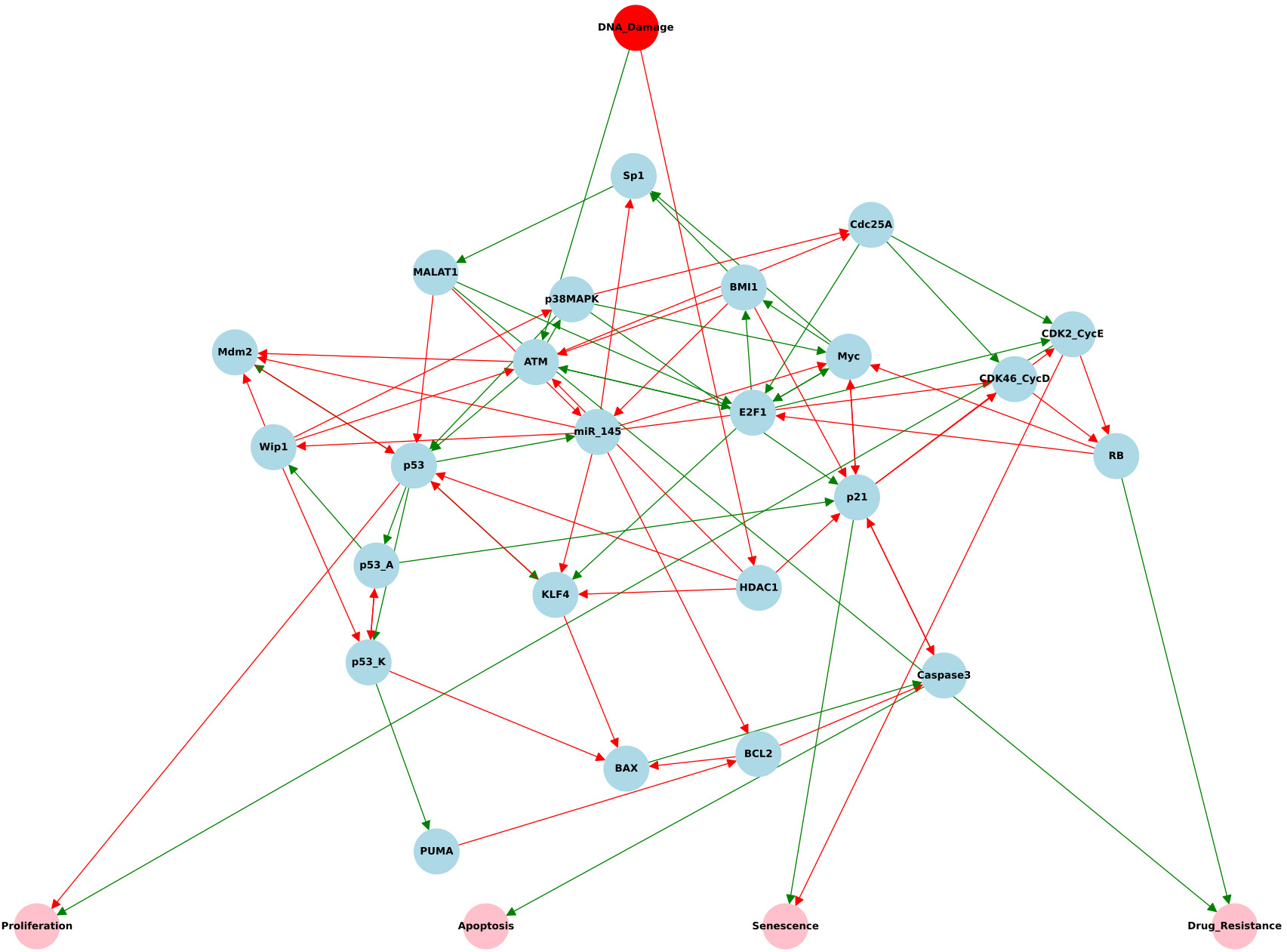}
	\caption{Reduced Boolean network for non-small cell lung cancer. For the construction of this network, the 31-node network was simplified with the most relevant genes and built based on the information in Table \ref{Atbl4}, resulting in a total of 29 nodes. Green arcs represent positive weights (activations), and red arcs represent negative weights (inhibitions). The red-colored node represents “DNA\_Damage” which corresponds to the input node; the four phenotypes that can be generated in this network are shown in pink, and the biological components corresponding to long non-coding RNAs, microRNAs, and various transcription factors are in blue.}
	\label{FIG:2}
\end{figure}

\subsection{Reduction to a 29-Node Network}

The main objective was to simplify the network while maintaining the essential dynamics that determine the key phenotypes. The first reduction stage involved removing the nodes Sirt\_1 and p53\_INP1. Although both nodes have relevant biological functions, their role within the global network dynamics was redundant. For instance, Sirt\_1 regulates p53 in a secondary manner, and its absence does not affect the final phenotype configurations. Similarly, p53\_INP1 acts as an aggregator of p53 activity, but this function is covered by other pathways within the network, as detailed further in section 4.2 of the discussion.

The reduced 29-node Boolean network maintained the four terminal phenotypes while simplifying its dynamical behavior by reducing the number of limit cycles to a single cycle of length 2. This network exhibited a total of $2^{25} = 33,554,432$ possible configurations, from which 50\% of the initial states converged to steady state 1 (proliferation), 49.89\% to steady state 2 (drug\_resistance), 0.01\% to steady state 3 (senescence), 0.04\% to steady state 4 (apoptosis), and 0.05\% to limit cycle 1. Importantly, this reduction preserved the behavior of key regulatory nodes, ensuring that the essential functional characteristics of the system remained intact. Furthermore, the attractor landscape of the reduced network was consistent with that of the original network, demonstrating that the reduction process did not alter the fundamental properties of state convergence. The results of this analysis are summarized in Table~\ref{tbl2}, while the structure of the reduced network is depicted in Figure~\ref{FIG:2}. For the simulation of this 29-node network, the rules of the 31-node network were adapted, and these new rules are shown in Table \ref{Atbl4}. 

\begin{table}[width=.9\linewidth,cols=7,pos=ht]
\caption{Steady state results for the 29-node network simulation.}
\label{tbl2}
\begin{tabular*}{\tblwidth}{@{} LLLLLLL@{} }
\toprule
Components & Steady state 1 & Steady state 2 & Steady state 3 & Steady state 4 & Limit Cycle 1\\
\midrule
ATM & 0 & 1 & 1 & 1 & 1 \ 1\\
p38MAPK & 0 & 1 & 1 & 1 & 1 \ 1\\
miR\_145 & 0 & 0 & 1 & 1 & 1 \ 1\\
Sp1 & 1 & 1 & 0 & 0 & 0 \ 0\\
MALAT1 & 1 & 1 & 0 & 0 & 0 \ 0\\
BMI1 & 1 & 1 & 0 & 0 & 0 \ 0\\
KLF4 & 1 & 1 & 0 & 0 & 0 \ 0\\
HDAC1 & 0 & 0 & 0 & 0 & 0 \ 0\\
Myc & 1 & 0 & 0 & 0 & 0 \ 0\\
p53 & 0 & 0 & 1 & 1 & 1 \ 1\\
Mdm2 & 1 & 0 & 0 & 0 & 0 \ 0\\
p53\_A & 0 & 0 & 0 & 1 & 1 \ 0\\
p53\_K & 0 & 0 & 1 & 0 & 1 \ 0\\
Wip1 & 0 & 0 & 0 & 0 & 0 \ 0\\
p21 & 0 & 0 & 1 & 1 & 0 \ 1\\
Cdc25A & 1 & 0 & 0 & 0 & 0 \ 0\\
CDK46\_CycD & 1 & 0 & 0 & 0 & 0 \ 0\\
CDK2\_CycE & 1 & 0 & 0 & 0 & 0 \ 0\\
RB & 0 & 1 & 1 & 1 & 1 \ 1\\
PUMA & 0 & 0 & 1 & 0 & 0 \ 1\\
BCL2 & 1 & 1 & 0 & 0 & 0 \ 0\\
BAX & 0 & 0 & 0 & 1 & 1 \ 0\\
E2F1 & 1 & 1 & 0 & 0 & 0 \ 0\\
Caspase3 & 0 & 0 & 0 & 1 & 0 \ 1\\
DNA\_Damage & 0 & 1 & 1 & 1 & 1 \ 1\\
Proliferation & 1 & 0 & 0 & 0 & - \ - \\
Drug Resistance & 0 & 1 & 0 & 0 & - \ - \\
Senescence & 0 & 0 & 1 & 1 & - \ - \\
Apoptosis & 0 & 0 & 0 & 1 & - \ - \\
\bottomrule
Attraction basin & 50.00 & 49.89 & 0.01 & 0.04 & 0.05\\
\end{tabular*}
\end{table}

\subsection{Reduction to a 14-Node Network}

In a second simplification stage, additional nodes were reduced whose configuration was fixed or deducible from other nodes in the context of activated “DNA\_Damage” (value 1). The resulting 14-node network, described in Figure \ref{FIG:A3}, continued to preserve the dynamics observed in the 31 and 29 node networks. In this case, the simulation led to three stable states and one limit cycle as shown in Table \ref{Atbl5}. The size of the configuration space was drastically reduced, facilitating a more efficient computational analysis.

For the simulation of this 14-node network, the rules of the 29-node network were adapted, and these new rules are shown in Table \ref{Atbl5}. This led to a total of $2^{14}=16,384$ possible configurations, with 98.44\% converging to steady state 1 (drug\_resistance), 0.39\% to steady state 2 (senescence), 0.39\% to steady state 3 (apoptosis), and 0.78\% to the limit cycle. These values are shown in Table \ref{Atbl6}. Upon analyzing the configurations of the nodes obtained and comparing them with the 31-node and 29-node networks, no changes in their values were observed, indicating that the dynamics remain unaffected.

\subsection{Reduction to a 9-Node Network}

\begin{figure}[ht]
	\centering
	\includegraphics[width=.375\textwidth]{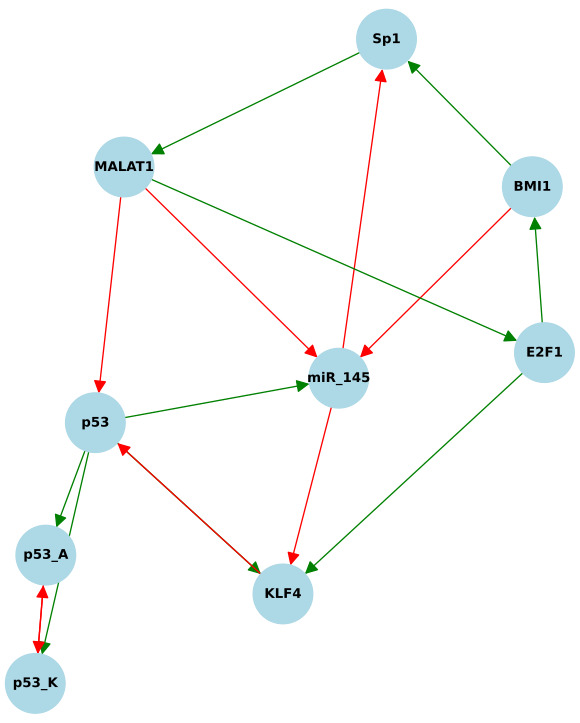}
	\caption{Minimal 9-node Boolean network for non-small cell lung cancer. To construct this network, the 14-node network was simplified from the information in Table \ref{Atbl7}, discarding nodes that can be deduced from their configurations by other nodes. Green arcs represent positive weights (activations), and red arcs represent negative weights (inhibitions).}
	\label{FIG:3}
\end{figure}

Finally, a minimally complex network comprising only 9 nodes was obtained, as shown in Figure \ref{FIG:3}. For this final reduction, all nodes exhibiting terminal‐node group behavior were removed. In Figure \ref{FIG:A3}, when identifying the nodes “p53\_K,” “KLF4,” “miR\_145,” “p53\_A,” and “BMI1,” these point to the group “BAX,” “Caspase3,” “p21,” “BCL2,” and “PUMA,” the latter exhibiting terminal‐node group behavior. These nodes were eliminated, and the rules were modified as presented in Table \ref{Atbl7}.

In the synchronous simulation, this yielded a total of $2^{9}=512$ possible configurations, of which 98.44\% converged to steady state 1 (drug resistance), 0.39\% to steady state 2 (senescence), 0.39\% to steady state 3 (apoptosis), and 0.78\% to the length two limit cycle. When comparing these outcomes with those from the 31, 29, and 14 node networks, no differences were observed, as shown in Table \ref{tbl3}. The steady state associated with the “Drug\_Resistance” phenotype exhibited a significantly larger attractor basin relative to the others; these basins of attraction are illustrated in Figure \ref{FIG:A4}

\begin{table}[width=.9\linewidth,cols=5,pos=ht]
\caption{Steady state results for the 9-node network simulation.}
\label{tbl3}
\begin{tabular*}{\tblwidth}{@{} LLLLLL@{} }
\toprule
Components & Steady state 1 & Steady state 2 & Steady state 3 & Limit Cycle 1\\
\midrule
miR\_145 & 0 & 1 & 1 & 1 \ 1 \\
Sp1 & 1 & 0 & 0 & 0 \ 0 \\
MALAT1 & 1 & 0 & 0 & 0 \ 0 \\
BMI1 & 1 & 0 & 0 & 0 \ 0 \\
KLF4 & 1 & 0 & 0 & 0 \ 0 \\
p53 & 0 & 1 & 1 & 1 \ 1 \\
p53\_A & 0 & 0 & 1 & 0 \ 1 \\
p53\_K & 0 & 1 & 0 & 0 \ 1 \\
E2F1 & 1 & 0 & 0 & 0 \ 0 \\
\bottomrule
Attraction basin & 98.44 & 0.39 & 0.39 & 0.78\\
\end{tabular*}
\end{table}

\subsection{Analysis of Update Schemes}

To evaluate the robustness of the reduced 9-node network, the method described in Section 2.4 of Materials and Methods was applied, identifying 10,632 representative deterministic update schemes. Among these, 7,356 (69.19\%) schemes exhibited only the three steady state, while the remaining 3,276 (30.81\%) included limit cycles. Of the schemes with cycles, 2,836 exhibited a single limit cycle, 362 included two limit cycles, and 78 showed five limit cycles.

The analysis of these schemes, summarized in Table \ref{Atbl8}, revealed that the three steady states were preserved in all cases, demonstrating the robustness of the network under different update dynamics. Regarding the limit cycles, all corresponded to two-state oscillations (cycles of length 2), with the most frequent one matching the cycle observed in the synchronous simulation, as detailed in Table \ref{Atbl9}.

\subsection{Rules Fitting}

In order to identify simpler or alternative rules that preserve the network’s expected dynamic behavior (that is, only the three steady states), we developed the Boolean Rule‐Fitting Algorithm. This optimized, exhaustive fitting method explores progressively larger combinations—ranging from one to three candidate regulatory nodes—one combination at a time, constructing Boolean expressions via the logical operators AND, OR, and NOT. Each candidate rule is evaluated in two stages: first, its local functionality is verified by checking whether it correctly reproduces the node’s value in the original attractors; then, if it passes this test, its global consistency is assessed by simulating the complete network dynamics under the new rule to confirm that the resulting attractors match exactly those desired. The full implementation is detailed in the Discussion, and the pseudocode is provided in the Appendix.

The algorithm generated five candidate rules that produced the target dynamics (that is, only the three steady states). After thorough review, only one rule proved biologically coherent within the context of the entire network: for the BMI1 node, the modified rule is “(!p53\_A \& !p53\_K) | E2F1”, Further justification for this selection is provided in the Discussion section. In the synchronous simulation, this yielded a total of $2^{9}=512$ possible configurations, of which 99.22\% converged to steady state 1 (drug resistance), 0.39\% to steady state 2 (senescence), and 0.39\% to steady state 3 (apoptosis). The results of this simulation are presented in Table \ref{Atbl11}.

As with the reduced 9-node network, the method described in Section 2.4 of Materials and Methods was applied, identifying 23,107 representative deterministic update schemes. Of these, 21,858 schemes (94.59\%) exhibited only the three steady states, while the remaining 1,243 schemes (5.41\%) included limit cycles. The analysis of these schemes, summarized in Table \ref{Atbl12}, revealed that the three steady states were preserved in all cases, demonstrating the robustness of the network under various update dynamics. Regarding the limit cycles, 586 schemes exhibited a single limit cycle, 534 included two limit cycles, and 129 showed three limit cycles. Furthermore, all detected cycles were of length 2, as detailed in Table \ref{Atbl13}.

\section{Discussion}

In Boolean network simulations, the configurations can reach steady states represented in this case as distinct phenotypes. Additionally, configurations can reach limit cycles, which are periodic oscillation patterns among a set of configurations. For example, a limit cycle of length 2 implies an alternation between two configurations. Configuration refers to the state of each node or biological component, indicating whether it is activated (1) or deactivated (0). For instance, each steady state or phenotype, as in Table \ref{tbl2}, has its own node states. A network may have multiple configurations. The total number of configurations is calculated by raising the number of possible states for each node (in this case, two for states 0 and 1) to the total number of nodes in the network. For example, a network with 3 nodes would have $2^{3} = 8$ possible configurations. These configurations can converge to a steady state or a limit cycle, which can be visualized in an attractor basin, as shown in Figure \ref{FIG:A4}. In this case, each colored dot represents a specific configuration of the network, with red dots representing configurations that converge to a single endpoint, in this case, the "Drug\_Resistance" phenotype. This occurs because, during the simulation, the configurations progressively evolve until reaching a steady state or a limit cycle.

\subsection{Importance of Complexity Reduction}

The main objective of this study is the reconstruction and reduction of Boolean networks, in this case applied to the study of drug resistance in NSCLC. The aim is to simplify the network structure without losing the essence of its dynamic behavior, in order to find the minimal network that maintains the initial dynamics, assess its robustness, and determine if it has biological significance as a bioinformatic tool for identifying candidate genes.

Once it was confirmed that the rules for each node have biological grounding, the first reduction of the network was carried out, with the goal of simplifying the structure while preserving the essential dynamics that determine key phenotypes. Therefore, an analysis in terms of biological functions was conducted, leading to the removal of the nodes Sirt\_1 and p53\_INP1. The function of Sirt\_1 is as a histone deacetylase that regulates various cellular processes, such as aging, inflammation, and stress response \cite{song2012janus}. In cancer, its function is ambivalent: on one hand, it can act as an oncogene by promoting cell proliferation and oxidative stress resistance in tumor cells; on the other hand, in certain contexts, it also acts as a tumor suppressor by reducing inflammation and protecting DNA from damage \cite{liu2009critical}. As for p53\_INP1 (Tumor protein p53 inducible nuclear protein 1), it is a stress-induced gene mainly regulated by p53. It promotes phosphorylation of p53 at serine 46, which increases its stability and transcriptional activity, crucial for activating genes such as p21 and PIG3, which halt cell growth and promote apoptosis in response to DNA damage \cite{shahbazi2013tumor}.

Although both Sirt\_1 and p53\_INP1 have important roles in cancer in general, their roles in this network are redundant. Reviewing the rules in Table 3 \ref{Atbl2}, Sirt\_1 is activated when E2F1 is active and miR\_145 and HDAC1 are inactive. Its main function is to regulate the nodes p53\_A and p53\_K, indirectly affecting p53, which influences network dynamics. However, p53 activation depends on multiple pathways, including ATM, KLF4, and Mdm2, which provide sufficient regulation to maintain the same steady states without the need for Sirt\_1. Since Sirt\_1 only has a secondary influence on p53 activation, its removal does not change the network's behavior regarding the activation of terminal phenotypes. For p53-INP1, it is a node activated when any form of p53 (p53\_A or p53\_K) is active, acting as an aggregator of p53 activity. Although p53-INP1 plays a role in activating p53\_A and regulating Wip1, this function is redundant, as p53 alone can activate p53\_A and other nodes. Additionally, Wip1 is also directly regulated by p53\_A, allowing Wip1’s function to be maintained without the need for p53-INP1. Therefore, the removal of p53-INP1 does not affect the network's ability to properly activate Wip1 or to maintain p53 regulation.

In this first elimination of these 2 nodes, the network was simplified, adapting the rules of the original 31-node network. These new rules are shown in Table \ref{Atbl4}. This resulted in a 29-node network that maintains the same steady states ("Proliferation," "Drug\_Resistance," "Senescence," and "Apoptosis") and reduces the number of limit cycles to one of length 2, without altering the final phenotypes, as shown in Table \ref{tbl2}. For this simulation, the terminal nodes were not considered, as they do not contribute to the overall dynamics; thus, 25 nodes were analyzed, and the phenotypes or terminal nodes were manually verified according to the rules. Comparing the configurations of each remaining node with the 31-node network showed no changes.

On the other hand, the network reduction shows that the nodes Sirt\_1 and p53\_INP1 sustain negative feedback loops (circuits with an odd number of inhibitions) capable of generating a limit cycle of period 4. Suppressing either of these nodes dismantles all negative cycles, causing the network dynamics to converge exclusively toward biologically relevant fixed points. In regulatory network theory, a negative cycle is a closed path in which the signal undergoes an odd number of negations, such that an initial increase in a node eventually feeds back as an inhibition upon itself after a full loop. This delayed self-inhibition can lead to sustained oscillations, provided that the internal delay is sufficient. This criterion stems from the “Thomas conditions” \cite{thomas1991regulatory}, which state that the existence of at least one negative circuit is a necessary (but not sufficient) condition for the emergence of periodic behavior, whereas positive circuits are associated with multistability. Therefore, the removal of Sirt\_1 or p53\_INP1 eliminates this structural source of oscillations, resulting in the disappearance of spurious limit cycles of length 4.

The reduction process carried out in this study demonstrates the feasibility of simplifying complex networks without compromising the essential dynamics that define the phenotypes observed in NSCLC. The removal of redundant nodes, such as Sirt\_1 and p53\_INP1, highlights the importance of identifying components that, although biologically relevant, do not significantly contribute to the final configurations of the network. This strategy not only facilitates computational analysis by reducing the configuration space but also can improve our understanding of the key interactions underlying cellular phenotypes. The consistency of steady states and the long two limit cycle in the 31, 29, 14, and 9 node networks validates the robustness of this approach.

\subsection{Robustness of Phenotypes Against Update Schemes}

The analysis of asynchronous update schemes in the minimal 9-node network provides additional evidence of model stability. The ability of the network to maintain the same three stable states and limit cycles under more than 10,000 representative schemes highlights the system's robustness to variations in update dynamics. This behavior reinforces the utility of the minimal network for reliably representing the mechanisms of drug resistance in NSCLC.

It is particularly notable that the "Drug Resistance" phenotype exhibits the largest attractor basin in both synchronous and asynchronous simulations. This result coherently reflects the clinical reality, where drug resistance is a prevalent phenomenon in NSCLC patients \cite{o2011role}. The congruence between computational simulations and clinical observations suggests that the remaining nodes in the minimal network adequately capture the main determinants of this phenotype.

For this new minimal 9-node network, which maintains the dynamics of the larger networks when “DNA Damage” is activated, we proceed to evaluate the stability of the network. If the network is robustly stable, it should maintain its dynamics when simulated with asynchronous update schemes, where the expected results are the consistent attainment of the 3 steady states and the appearance of the same or different limit cycles.

In Boolean networks, a synchronous simulation means that all network nodes update their states simultaneously at each time step, allowing observation of a uniform dynamic that facilitates the identification of steady states or limit cycles. This reflects an idealized view of the system, where all cellular processes would respond in a coordinated manner. This approach has been applied to all analyses conducted so far. In contrast, in asynchronous simulation, nodes update their states independently, introducing variability in the activation timing of each component. This type of simulation allows for capturing a more flexible and realistic dynamic, as biological systems often operate non-synchronously \cite{das2015inference} \cite{saadatpour2010attractor}. This approach can reveal additional behaviors, such as increases in the number of cycles, that are not observed in synchronous simulation.

\subsection{Biological Relevance of Minimal Network Components}

In analyzing the roles of these 9 nodes, a search was conducted for each component of the minimal network to understand their associations with NSCLC and cancer. It was found that:

\begin{itemize}
\item miR-145 (microRNA-145-5p): This microRNA acts as a tumor suppressor in NSCLC by regulating cell migration and invasion. Its decreased expression is associated with increased chemotherapy resistance due to interaction with Sp1 and BMI1, both related to the epithelial-mesenchymal transition process \cite{chang2022mir}\cite{shen2015low}.

\item BMI1 (B lymphoma Mo-MLV insertion region 1 homolog): As part of the PRC1 complex, BMI1 is associated with epigenetic regulation in NSCLC and promotes tumor cell invasion and proliferation by inhibiting the expression of tumor suppressor genes. High BMI1 expression is associated with a worse prognosis in NSCLC patients \cite{zhang2019lncrna}.

\item Sp1 (Specificity protein 1): This transcription factor is involved in regulating key genes for cell proliferation and has been shown to collaborate with BMI1 to induce treatment resistance in NSCLC. Inhibiting Sp1 could help restore chemotherapy sensitivity \cite{chang2022mir}.

\item MALAT1 (Metastasis-associated lung adenocarcinoma transcript 1): This long non-coding RNA (lncRNA) is implicated in promoting tumor growth in NSCLC by inhibiting microRNA-613 and increasing the expression of COMMD8, which promotes cell proliferation and reduces apoptosis. Its activity is therefore crucial in regulating tumor progression \cite{wang2020lncrna}.

\item E2F1 (E2F transcription factor 1): This transcription factor regulates genes involved in cell cycle progression. Its inappropriate activation is associated with increased proliferation and cell migration in NSCLC, contributing to tumor aggressiveness \cite{donzelli2012microrna}.

\item p53 (Tumor suppressor p53 protein): This tumor suppressor gene is essential in regulating the cell cycle and apoptosis. Its inactivation or mutation facilitates tumor proliferation in NSCLC, and increasing its activity through p53 restoration could inhibit tumor growth \cite{quan2019p53} \cite{wang2013mir}.

\item p53\_A (Tumor suppressor p53 protein Ser-15 and Ser-20): is crucial for its activation in response to DNA damage, facilitated by the ATM kinase, promoting the tumor suppressor function of p53 by regulating genes that induce apoptosis \cite{saito2002atm}. Phosphorylation of p53 at Ser-15 and Ser-20 in A549 lung cancer cells exposed to cisplatin and air pollutants demonstrates that these sites are specifically activated in response to DNA damage, contributing to cell cycle arrest and controlled cell damage \cite{niechoda2023cell}.

\item p53\_K (Tumor suppressor p53 protein Ser-46): Inhibition of p53 phosphorylation at Ser-46, induced by the drug rapamycin, has been shown to reduce its apoptotic activity in A549 lung cancer cells exposed to actinomycin D. This suggests that phosphorylation at Ser-46 is crucial for maintaining p53’s role in DNA damage-induced apoptosis \cite{krzesniak2014rapamycin}.

\item KLF4 (Krüppel-like Factor 4): KLF4 is a transcription factor that acts as a tumor suppressor and regulates cell proliferation and differentiation in NSCLC. Its reduced expression, promoted by specific microRNAs such as miR-3120-5p, is associated with increased metastatic potential \cite{xu2018mir}. Although it is active in the case of "Drug\_resistance," it depends on various genes and proteins, including p21, to regulate its function in cancer. KLF4 regulates this cell cycle inhibitor, which is essential for growth control and cell cycle progression, especially in its role as a tumor suppressor in different cancer contexts \cite{wei2010klf4alpha} \cite{rowland2006klf4}. Observing the state of p21 in the 31-, 29-, and 14-node networks, in the "Drug Resistance" phenotype, it is inactive, indicating biological concordance of these nodes in this minimal network in generating the 3 phenotypes. Thus, this minimal network reflects the behavior of these genes concerning their phenotypes in NSCLC.

\end{itemize}

\subsection{Boolean Rule Fitting for Enhanced Network Dynamics}

In the context of Boolean networks applied to biology, an attractor is a set of states toward which a dynamical system tends, commonly representing stable cellular states such as differentiated cell types. These attractors are fundamental for modeling the stable behavior of cells and their fate in processes such as differentiation, proliferation, or apoptosis \cite{kauffman1969metabolic} \cite{thomas1973boolean}. However, in complex computational models, what some authors conceptualize as spurious attractors may arise—these are steady states or limit cycles that do not correspond to real biological behaviors \cite{kervizic2008dynamical}. Such attractors are often the result of model limitations, poorly defined transition functions, or noisy or incomplete input data. These configurations can significantly affect the interpretation of the model, leading to false predictions about the system's stability or its response to perturbations. To mitigate this problem, approaches such as the constrained genetic algorithm (CGA-BNI) have been developed; using steady-state gene-expression data, it imposes path-consistency constraints and selects Boolean networks whose attractors more faithfully match the observed states, demonstrating a significant improvement in both dynamic and structural accuracy compared with other existing methods \cite{trinh2021novel}.

In our case, we developed an exhaustive and optimized fitting method aimed at generating and validating alternative Boolean rules in a gene regulatory network, with the goal of preserving only the attractors of interest. Starting from an initial set of rules, the method simulates the network's steady states and then replaces each rule with new logical combinations constructed from other genes. For each gene, it evaluates whether these candidate rules are functionally equivalent in the steady states and, most importantly, whether they reproduce exactly the same set of desired attractors and not spuriou ones. In this specific case, the original network produces three steady states and one spuriou limit cycle (i.e., mixing different types of undesired configurations). Therefore, the code is designed to identify an alternative dynamic that eliminates the spuriou cycle and retains only the three relevant attractors.

The Boolean Rule Fitting Algorithm operates under the principle that for each target node, the algorithm explores increasingly complex combinations of possible regulatory nodes, generating candidate Boolean expressions using logical operators (AND, OR, NOT). Each candidate rule is evaluated on two levels: first, its local functionality is checked, that is, whether it correctly reproduces the value of the node in the original attractors. If it passes this test, its global consistency is then assessed by simulating the full network dynamics under the new rule to determine whether the resulting attractors match exactly the desired ones.

To generate rules, logical combinations were explored that involve one, two or three potential regulators of the target node, applying a construction logic based on standard Boolean operators: negation (!), conjunction (\&), and disjunction (|). In the case of a single regulatory variable, the set of candidate expressions was limited to two possible forms: the variable in its direct form and its negation. For two variables, all possible combinations were generated using the mentioned operators, including variants with one or both variables negated. This yielded a total of eight configurations per pair, such as [X \& Y], [!X \& Y], [X | !Y], [!X | !Y], etc., covering a basic yet expressive range of binary logical relationships.

For combinations involving three regulators, a broader strategy was adopted that included expressions without parentheses and expressions with explicit parentheses. The parenthesis-free expressions consisted of direct combinations of the three terms, evaluated according to standard logical precedence (where "!" has higher priority than "\&", and "\&" has higher priority than "|"). Typical examples include expressions such as [X \& Y \& Z], [!X | Y | Z], [!X \& !Y \& Z], etc. Although syntactically simple, these forms capture multiple modes of simultaneous activation and repression among regulators.

On the other hand, parenthesized expressions were used to represent more complex hierarchical or conditional interactions, where two inputs are logically grouped before interacting with a third. Examples of these structures include [(!X \& Y) | Z], [(X | !Y) \& !Z], [(X \& Y) | !Z], etc. This type of construction reflects common patterns in gene regulation, where certain effects manifest only under specific combinations of signals, adding realism to the logical model.

This approach enables exploration of a broad expressive space while maintaining an interpretable and biologically plausible logical structure. The diversity of generated forms ensures that both simple rules and complex interactions are considered, thereby facilitating the identification of alternative rules that preserve the desired dynamics of the network while eliminating spuriou behaviors inconsistent with the biological framework. Rule generation was limited to combinations of up to three regulators, according to the Kauffman K parameter, which was set to a maximum of 3.

Kauffman's Boolean network parameter K controls connectivity and has a crucial impact on system behavior: When K is low, the system is ordered; when K is high, it becomes chaotic. Therefore, there exists a critical value Kc that marks the transition between order and chaos.

Classical studies, such as those by Derrida \cite{derrida1986random}, estimated this critical value at Kc=2 for unbiased Boolean functions. However, investigations like those by Zertuche have corrected this value by taking into account Boolean irreducibility, a structural property of the functions used. With this correction, the critical value is adjusted to \(K_c \approx 2.62\) \cite{zertuche2014phase}.

Networks with K > 3 are in the chaotic phase, characterized by high sensitivity to initial conditions, a large number of attractors, and loss of system stability. This behavior has been confirmed through simulations and analyses in different network topologies, including scale-free networks and generalized models \cite{iguchi2005rugged} \cite{sole1995phase}.

After running the fitting algorithm, five candidate rules were identified, from which the modified BMI1 variant defined as "(!p53\_A \& !p53\_K) | E2F1" was selected and grounded in the most recent experimental evidence. Ni et al. (2022) and Shi et al. (2024) demonstrated that, in non-small cell lung carcinoma (NSCLC), E2F1 is overexpressed, binds directly to the BMI1 promoter, and enhances its transcription, thereby promoting epithelial–mesenchymal transition (EMT), invasion, and chemoresistance \cite{ni2022disulfiram} \cite{yan2024angiogenesis}. In contrast, Parfenyev et al. (2021) and Wang et al. (2022) have shown that functional p53 (when phosphorylated at Ser15/Ser20 or Ser46) induces miR-200c and other microRNA suppressors that post-transcriptionally repress BMI1; loss of these phosphorylations attenuates miR-200c transcription, permitting BMI1 accumulation and fostering tumor plasticity and drug resistance \cite{parfenyev2021interplay} \cite{guo2025microrna}. Thus, the logical conjunction that incorporates E2F1 as an activator and p53 as a repressor more faithfully captures the dynamic balance between proliferative signals and tumor-suppressive mechanisms, aligning the Boolean model with empirical observations.

Subsequently, after simulating the modified rule and confirming that the original steady states remained unchanged, this adjustment was applied to the 29-node network. As in the reduced nine-node version, the length-two limit cycle was abolished and the steady-state configurations remained intact. Finally, upon evaluating asynchronous update schemes and comparing the 9 node network before and after this modification, the proportion of schemes yielding exclusively steady states increased from 69.19\% to 94.59\%, thereby markedly reducing the number of schemes that produced spuriou cycles.

\subsection{Therapeutic Implications}

Analyzing in detail, the largest attractor basin for both synchronous and asynchronous simulation corresponds to "Drug Resistance," covering 98.43\% of the total configurations leading to this phenotype in the synchronous simulation for the network without fitting; in the case of the modified 9-node network, this value corresponds to 99.22\%. While in the asynchronous simulation, it covers approximately 85\% for the network without fitting, in the case of the modified network, this value corresponds to 90\%. The attractor basin size suggests that the configuration with the highest likelihood of occurrence is "Drug Resistance." Clinical studies, such as \citep{o2011role}, indicate that 20\% to 30\% of NSCLC patients show objective responses to the Cisplatin and Pemetrexed combination \cite{o2011role}, drugs considered in constructing the original network used for this study. This implies significant resistance in most patients treated with this combination. This percentage indicates that approximately 70\% to 80\% of patients do not respond favorably to this treatment, highlighting the considerable challenge that drug resistance poses in managing NSCLC. The results suggest the potential use of these 9 genetic components as biological markers for "Drug Resistance" regarding the use of Cisplatin and Pemetrexed in treating non-small cell lung cancer or the use of these components in gene therapy promoting senescence and apoptosis states for this cancer.

The identification of attractor configurations associated with undesirable phenotypes, such as drug resistance, also opens the possibility of designing genetic or pharmacological interventions aimed at modifying these configurations. In this sense, the simplified model could serve as a useful tool for exploring therapeutic strategies computationally before experimental validation. Rules that satisfy both criteria are considered valid. At the end of the process, an alternative set of rules is generated for each node, reproducing the expected dynamics without introducing spuriou behaviors.

\section{Conclusions}

Our results demonstrate that aggressive down-scaling of a biologically grounded Boolean network can be achieved without sacrificing dynamical fidelity. The 9-node core retains all significant steady states and asynchronous representative update schemes, underscoring its topological and dynamic robustness. This simplification distills a small set of master regulators—miR-145, MALAT1, BMI1, Sp1, KLF4, E2F1 and phosphorylated p53 isoforms—that collectively govern proliferation, senescence and drug-resistance phenotypes in NSCLC. The resulting model is therefore well suited for in-silico screening of perturbations and for guiding focused wet-lab experiments. While the present findings are based solely on computational evidence, they chart a clear path toward experimentally testable interventions aimed at overcoming chemoresistance in lung cancer.

\clearpage
\renewcommand{\thepage}{\arabic{page}}

\section{Appendices}

\renewcommand{\thetable}{A.\arabic{table}}
\setcounter{table}{0} 

\renewcommand{\thefigure}{A.\arabic{figure}}
\setcounter{figure}{0} 


\begin{table}[width=.9\linewidth,cols=4,pos=ht]
\caption{All the distinct dynamics (seen as transition tables) that a BN with the digraph $G$ and the Boolean local functions given in the Example \ref{ex_1} can have by using the (representative) schemes $s_1 $, ..., $s_9$.}
\label{Atbl1}
\begin{tabular*}{\tblwidth}{@{} LLLLLLLLLL@{} }
\toprule
State & Sched. 1 & Sched. 2 & Sched. 3 & Sched. 4 & Sched. 5 & Sched. 6 & Sched. 7 & Sched. 8 & Sched. 9 \\
\midrule
& $s_{1}$(A)=1 & $s_{2}$(A)=2 & $s_{3}$(A)=1 & $s_{4}$(A)=2 & $s_{5}$(A)=3 & $s_{6}$(A)=2 & $s_{7}$(A)=1 & $s_{8}$(A)=1 & $s_{9}$(A)=1 \\
& $s_{1}$(B)=1 & $s_{2}$(B)=1 & $s_{3}$(B)=2 & $s_{4}$(B)=1 & $s_{5}$(B)=1 & $s_{6}$(B)=2 & $s_{7}$(B)=2 & $s_{8}$(B)=1 & $s_{9}$(B)=3 \\
& $s_{1}$(C)=1 & $s_{2}$(C)=2 & $s_{3}$(C)=1 & $s_{4}$(C)=1 & $s_{5}$(C)=2 & $s_{6}$(C)=1 & $s_{7}$(C)=2 & $s_{8}$(C)=2 & $s_{9}$(C)=2 \\
\midrule
000 & 000 & 000 & 000 & 000 & 000 & 000 & 000 & 000 & 000 \\
001 & 110 & 110 & 100 & 010 & 010 & 000 & 110 & 111 & 100 \\
010 & 000 & 000 & 000 & 000 & 000 & 000 & 000 & 000 & 000 \\
011 & 110 & 110 & 100 & 010 & 010 & 000 & 111 & 111 & 111 \\
100 & 000 & 000 & 000 & 000 & 000 & 000 & 000 & 000 & 000 \\
101 & 110 & 111 & 100 & 010 & 111 & 000 & 110 & 111 & 100 \\
110 & 001 & 000 & 011 & 101 & 000 & 111 & 000 & 000 & 000 \\
111 & 111 & 111 & 111 & 111 & 111 & 111 & 111 & 111 & 111 \\
\bottomrule
\end{tabular*}
\end{table}


\begin{figure}[ht]
	\centering
	\includegraphics[width=1\textwidth]{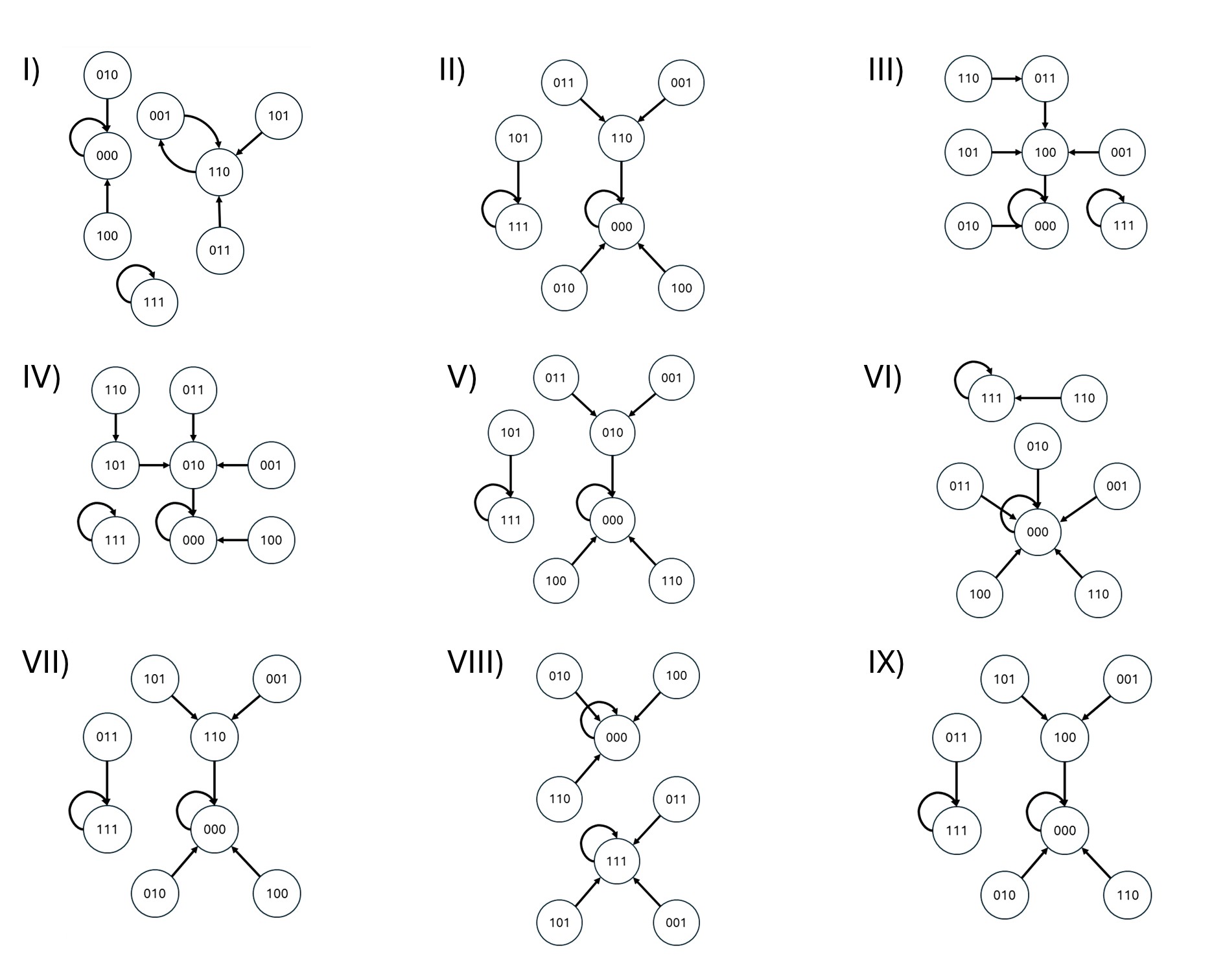}
	\caption{All the distinct dynamics of Table \ref{Atbl1} seen as state transition graphs for the Example \ref{ex_1}, where I), ..., IX) are associated to the (representative) schemes $s_1 $, ..., $s_9$, respectively.}
	\label{FIG:A1}
\end{figure}


\begin{figure}[ht]
	\centering
	\includegraphics[width=1\textwidth]{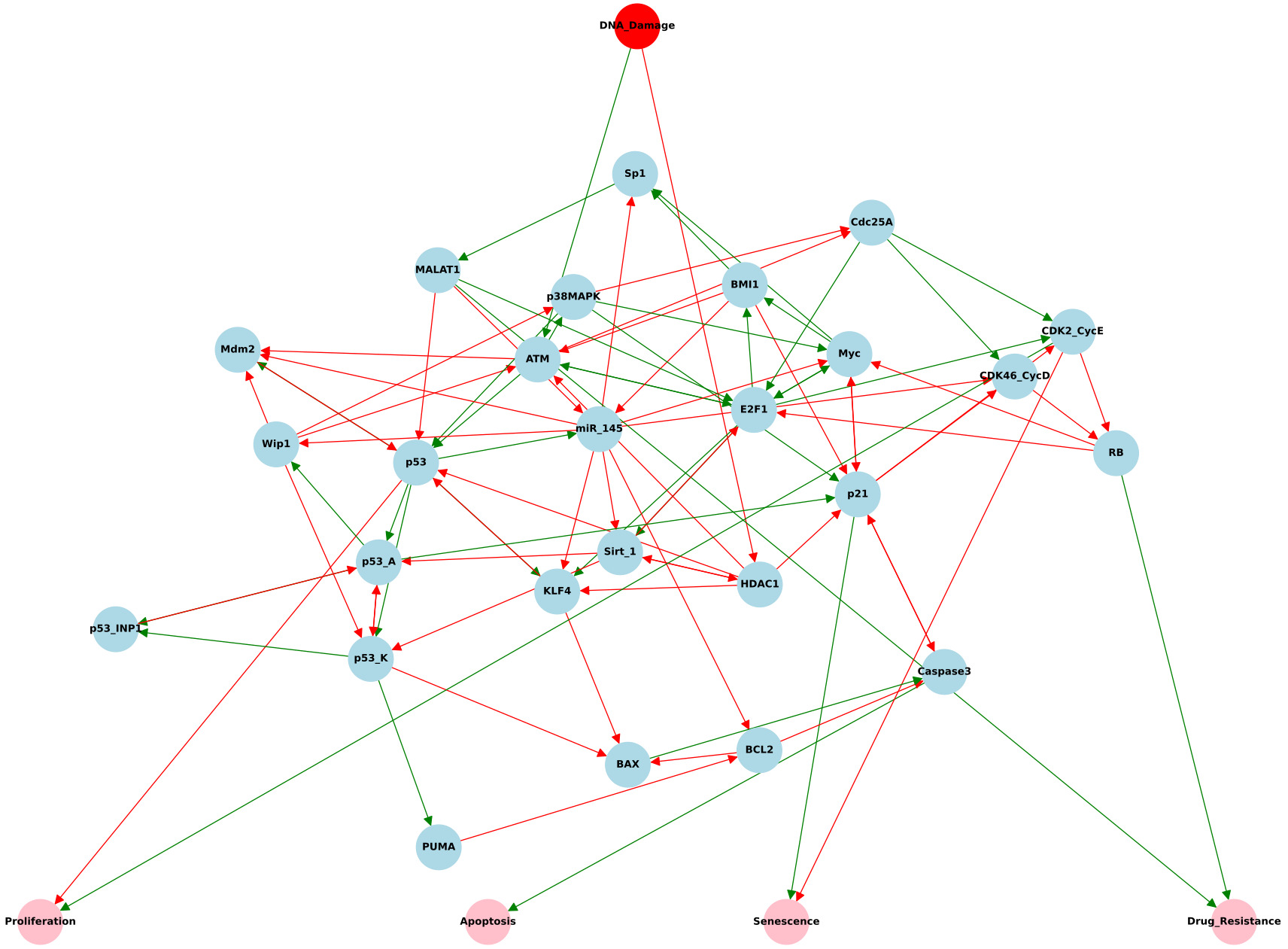}
	\caption{The 31-node Boolean network for non-small cell lung cancer reconstructed from the one that appears in \citep{gupta2024dynamic}, where green arcs represent positive weights (activations), and red arcs represent negative weights (inhibitions). The red-colored node represents "DNA\_Damage", which corresponds to the input node; the four phenotypes that can be generated in this network are shown in pink, and the biological components corresponding to long non-coding RNAs, microRNAs, and various transcription factors are in blue.}
	\label{FIG:A2}
\end{figure}


\begin{table}[width=.9\linewidth,cols=6,pos=ht]
\caption{The 31 local boolean functions (or interactions) associated to the nodes of the NSCLC network shown in Figure \ref{FIG:A2} (the input node DNA\_Damage is omitted).}
\label{Atbl2}
\begin{tabular*}{\tblwidth}{@{} LL@{} }
\toprule
Node & Interactions\\
\midrule
ATM & DNA\_Damage \& (!HDAC1 | !Wip1 | E2F1 | !BMI1)\\
p38MAPK & ATM \& !Wip1\\
miR\_145 & p53 \& !MALAT1 \& !BMI1\\
Sp1 & (BMI1 \& Myc) | !miR\_145\\
MALAT1 & Sp1\\
BMI1 & Myc | E2F1\\
KLF4 & !miR\_145 | (E2F1 \& !HDAC1 \& p53)\\
HDAC1 & !Sirt\_1 \& !DNA\_Damage\\
Myc & (E2F1 | p38MAPK | !p21) \& !RB \& !miR\_145\\
p53 & (ATM \& !KLF4) | (!Mdm2 \& p38MAPK \& !HDAC1 \& !MALAT1)\\
Mdm2 & (!Wip1 | p53) \& !ATM \& !miR\_145\\
p53\_A & !Sirt\_1 \& !p53\_K \& (p53 | !p53\_INP1)\\
p53\_K & !p53\_A \& (!Sirt\_1 | !Wip1) \& p53\\
Sirt\_1 & E2F1 \& !miR\_145 \& !HDAC1\\
p53\_INP1 & p53\_A | p53\_K\\
Wip1 & p53\_A \& !miR\_145\\
p21 & p53\_A | ((!HDAC1 \& !Myc) \& !BMI1 \& !Caspase3 \& p38MAPK)\\
Cdc25A & !ATM \& !p38MAPK\\
CDK46\_CycD & Cdc25A \& !miR\_145 \& !p21\\
CDK2\_CycE & Cdc25A \& E2F1 \& !p21\\
RB & !CDK46\_CycD \& !CDK2\_CycE\\
PUMA & p53\_K\\
BCL2 & !PUMA \& !miR\_145\\
BAX & (!BCL2 | !KLF4) \& !p53\_K\\
E2F1 & (!RB \& ((Cdc25A \& ATM) | !Sirt\_1)) | MALAT1 | Myc\\
Caspase3 & (!BCL2 | !p21) \& BAX\\
Proliferation & CDK2\_CycE \& !p53\\
Drug\_Resistance & MALAT1 \& RB\\
Senescence & p21 \& !CDK2\_CycE\\
Apoptosis & Caspase3\\
\bottomrule
\end{tabular*}
\end{table}


\begin{table}[width=.9\linewidth,cols=7,pos=ht]
\caption{Steady state results for the 31-node network simulation.}
\label{Atbl3}
\begin{tabular*}{\tblwidth}{@{}LLLLLLLL@{} }
\toprule
Components & Steady state 1 & Steady state 2 & Steady state 3 & Steady state 4 & Limit Cycle 1 & Limit Cycle 2 & Limit Cycle 3\\
\midrule
ATM & 0 & 1 & 1 & 1 & 1 1 & 0 0 0 0 & 0 0 0 0 \\ 
p38MAPK & 0 & 1 & 1 & 1 & 1 1 & 0 0 0 0 & 0 0 0 0 \\ 
miR\_145 & 0 & 0 & 1 & 1 & 1 1 & 0 0 0 0 & 0 0 0 0 \\ 
Sp1 & 1 & 1 & 0 & 0 & 0 0 & 1 1 1 1 & 1 1 1 1 \\ 
MALAT1 & 1 & 1 & 0 & 0 & 0 0 & 1 1 1 1 & 1 1 1 1 \\ 
BMI1 & 1 & 1 & 0 & 0 & 0 0 & 1 1 1 1 & 1 1 1 1 \\ 
KLF4 & 1 & 1 & 0 & 0 & 0 0 & 1 1 1 1 & 1 1 1 1 \\ 
HDAC1 & 0 & 0 & 0 & 0 & 0 0 & 1 1 1 1 & 1 0 1 0 \\ 
Myc & 1 & 0 & 0 & 0 & 0 0 & 1 1 0 0 & 1 1 0 1 \\ 
p53 & 0 & 0 & 1 & 1 & 1 1 & 0 0 0 0 & 0 0 0 0 \\ 
Mdm2 & 1 & 0 & 0 & 0 & 0 0 & 0 0 1 1 & 0 1 1 1 \\ 
p53\_A & 0 & 0 & 0 & 1 & 1 0 & 0 0 1 1 & 0 0 1 0 \\ 
p53\_K & 0 & 0 & 1 & 0 & 1 0 & 0 0 0 0 & 0 0 0 0 \\ 
Sirt\_1 & 1 & 1 & 0 & 0 & 0 0 & 0 0 0 0 & 1 0 1 0 \\ 
p53\_INP1 & 0 & 0 & 1 & 1 & 0 1 & 1 0 0 1 & 0 0 0 1 \\ 
Wip1 & 0 & 0 & 0 & 0 & 0 0 & 1 0 0 1 & 0 0 0 1 \\ 
p21 & 0 & 0 & 1 & 1 & 0 1 & 1 0 0 1 & 0 0 0 1 \\ 
Cdc25A & 1 & 0 & 0 & 0 & 0 0 & 1 1 1 1 & 1 1 1 1 \\ 
CDK46\_CycD & 1 & 0 & 0 & 0 & 0 0 & 0 0 1 1 & 0 1 1 1 \\ 
CDK2\_CycE & 1 & 0 & 0 & 0 & 0 0 & 0 0 1 1 & 0 1 1 1 \\ 
RB & 0 & 1 & 1 & 1 & 1 1 & 0 1 1 0 & 0 1 0 0 \\ 
PUMA & 0 & 0 & 1 & 0 & 0 1 & 0 0 0 0 & 0 0 0 0 \\ 
BCL2 & 1 & 1 & 0 & 0 & 0 0 & 1 1 1 1 & 1 1 1 1 \\ 
BAX & 0 & 0 & 0 & 1 & 1 0 & 0 0 0 0 & 0 0 0 0 \\ 
E2F1 & 1 & 1 & 0 & 0 & 0 0 & 1 1 1 1 & 1 1 1 1 \\ 
Caspase3 & 0 & 0 & 0 & 1 & 0 1 & 0 0 0 0 & 0 0 0 0 \\ 
DNA\_Damage & 0 & 1 & 1 & 1 & 1 1 & 0 0 0 0 & 0 0 0 0 \\
Proliferation & 1 & 0 & 0 & 0 & - \ -  & - \ - \ - \ -  &  - \ - \ - \ - \\
Drug Resistance & 0 & 1 & 0 & 0 & - \ -  &  - \ - \ - \ -  &  - \ - \ - \ - \\
Senescence & 0 & 0 & 1 & 1 & - \ -  & - \ - \ - \ -  &  - \ - \ - \ -  \\
Apoptosis & 0 & 0 & 0 & 1 & - \ -  &  - \ - \ - \ -  &  - \ - \ - \ -  \\
\bottomrule
\end{tabular*}
\end{table}


\begin{table}[width=.9\linewidth,cols=6,pos=ht]
\caption{Adapted rules for the 29-node network.}
\label{Atbl4}
\begin{tabular*}{\tblwidth}{@{} LL@{} }
\toprule
Node & Interactions\\
\midrule
ATM & DNA\_Damage \& (!HDAC1 | !Wip1 | E2F1 | !BMI1) \\
p38MAPK & ATM \& !Wip1 \\
miR\_145 & p53 \& !MALAT1 \& !BMI1 \\
Sp1 & (BMI1 \& Myc) | !miR\_145 \\
MALAT1 & Sp1 \\
BMI1 & Myc | E2F1 \\
KLF4 & !miR\_145 | (E2F1 \& !HDAC1 \& p53) \\
HDAC1 & !DNA\_Damage \\
Myc & (E2F1 | p38MAPK | !p21) \& !RB \& !miR\_145 \\
p53 & (ATM \& !KLF4) | (!Mdm2 \& p38MAPK \& !HDAC1 \& !MALAT1) \\
Mdm2 & (!Wip1 | p53) \& !ATM \& !miR\_145 \\
p53\_A & !p53\_K \&  (p53) \\
p53\_K & !p53\_A \& (!Wip1) \& p53 \\
Wip1 & p53\_A  \& !miR\_145 \\
p21 & p53\_A | ((!HDAC1 \& !Myc) \& !BMI1 \& !Caspase3 \& p38MAPK) \\
Cdc25A & !ATM \& !p38MAPK \\
CDK46\_CycD & Cdc25A \& !miR\_145 \& !p21 \\
CDK2\_CycE & Cdc25A \& E2F1 \& !p21 \\
RB & !CDK46\_CycD \& !CDK2\_CycE \\
PUMA & p53\_K \\
BCL2 & !PUMA \& !miR\_145 \\
BAX & (!BCL2 | !KLF4) \& !p53\_K \\
E2F1 & (!RB \& ((Cdc25A \& ATM))) | MALAT1 | Myc \\
Caspase3 & (!BCL2 | !p21) \& BAX \\
Proliferation & CDK2\_CycE \& !p53 \\
Drug\_Resistance & MALAT1 \& RB \\
Senescence & p21 \& !CDK2\_CycE \\
Apoptosis & Caspase3 \\
\bottomrule
\end{tabular*}
\end{table}


\begin{figure}[ht]
	\centering
	\includegraphics[width=.5\textwidth]{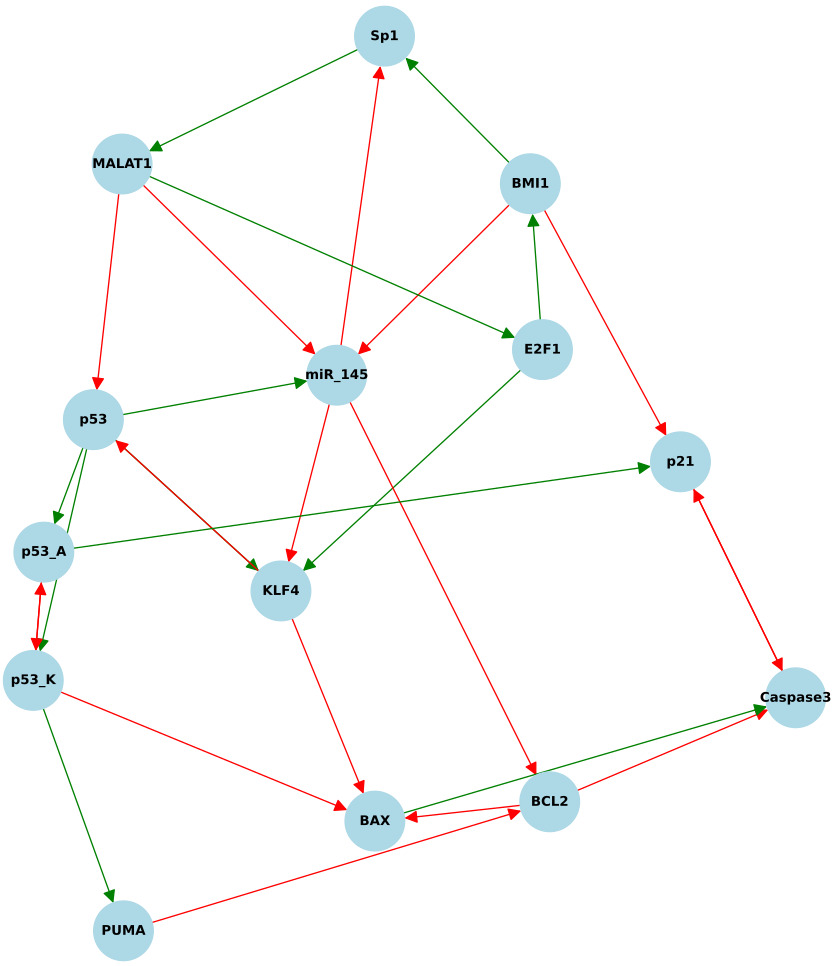}
	\caption{Reduced 14-node Boolean network for non-small cell lung cancer. To construct this network, the 29-node network was simplified from the information in Table \ref{Atbl5}. The 29-node network was simplified for the case where “DNA\_Damage” equals 1, discarding nodes that fixed their configuration similarly in steady states and limit cycles, and also eliminating terminal nodes. Green arcs represent positive weights (activations), and red arcs represent negative weights (inhibitions).}
	\label{FIG:A3}
\end{figure}


\begin{table}[width=.9\linewidth,cols=6,pos=ht]
\caption{Adapted rules for the 14-node network.}
\label{Atbl5}
\begin{tabular*}{\tblwidth}{@{} LL@{} }
\toprule
Node & Interactions\\
\midrule
miR\_145 & p53 \& !MALAT1 \& !BMI1 \\
Sp1 & (BMI1) | !miR\_145 \\
MALAT1 & Sp1 \\
BMI1 & E2F1 \\
KLF4 & !miR\_145 | (E2F1 \& p53) \\
p53 & !KLF4 | !MALAT1 \\
p53\_A & !p53\_K \& p53 \\
p53\_K & !p53\_A \& p53 \\
BAX & (!BCL2 | !KLF4) \& !p53\_K \\
E2F1 & MALAT1 \\
Caspase3 & (!BCL2 | !p21) \& BAX \\
p21 & p53\_A | (!BMI1 \& !Caspase3) \\
BCL2 & !PUMA \& !miR\_145 \\
PUMA & p53\_K \\
\bottomrule
\end{tabular*}
\end{table}


\begin{table}[width=.9\linewidth,cols=6,pos=ht]
\caption{Steady state results for the 14-node network simulation.}
\label{Atbl6}
\begin{tabular*}{\tblwidth}{@{} LLLLLL@{} }
\toprule
Components & Steady state 1 & Steady state 2 & Steady state 3 & Limit Cycle 1\\
\midrule
miR\_145 & 0 & 1 & 1 & 1 \ 1\\
Sp1 & 1 & 0 & 0 & 0 \ 0\\
MALAT1 & 1 & 0 & 0 & 0 \ 0\\
BMI1 & 1 & 0 & 0 & 0 \ 0\\
KLF4 & 1 & 0 & 0 & 0 \ 0\\
p53 & 0 & 1 & 1 & 1 \ 1\\
p53\_A & 0 & 0 & 1 & 1 \ 0\\
p53\_K & 0 & 1 & 0 & 1 \ 0\\
p21 & 0 & 1 & 1 & 0 \ 1\\
PUMA & 0 & 1 & 0 & 0 \ 1\\
BCL2 & 1 & 0 & 0 & 0 \ 0\\
BAX & 0 & 0 & 1 & 1 \ 0\\
E2F1 & 1 & 0 & 0 & 0 \ 0\\
Caspase3 & 0 & 0 & 1 & 0 \ 1\\
\bottomrule
Attraction basin & 98.44 & 0.39 & 0.39 & 0.78\\
\end{tabular*}
\end{table}


\begin{table}[width=.9\linewidth,cols=6,pos=ht]
\caption{Adapted rules for the 9-node network.}
\label{Atbl7}
\begin{tabular*}{\tblwidth}{@{} LL@{} }
\toprule
Node & Interactions\\
\midrule
miR\_145 & p53 \& !MALAT1 \& !BMI1 \\
Sp1 & (BMI1) | !miR\_145 \\
MALAT1 & Sp1 \\
BMI1 & E2F1 \\
KLF4 & !miR\_145 | (E2F1 \& p53) \\
p53 & !KLF4 | !MALAT1 \\
p53\_A & !p53\_K \& p53 \\
p53\_K & !p53\_A \& p53 \\
E2F1 & MALAT1 \\
\bottomrule
\end{tabular*}
\end{table}


\begin{table}[width=.9\linewidth,cols=4,pos=ht]
\caption{Limit cycle analysis for asynchronous update schemes of the minimum 9-node network.}
\label{Atbl8}
\begin{tabular*}{\tblwidth}{@{} LLLL@{} }
\toprule
Steady state Configuration & Average Basin Attraction & SD & Count \\
\midrule
011110001 & 441.19 & 69.43 & 10,632 \\
100001010 & 23.25 & 21.86 & 10,632 \\
100001100 & 23.25 & 21.86 & 10,632 \\
\bottomrule
\end{tabular*}
\end{table}


\begin{table}[width=.9\linewidth,cols=4,pos=ht]
\caption{Top 10 limit cycle for asynchronous update schemes for the minimum 9-node network.}
\label{Atbl9}
\begin{tabular*}{\tblwidth}{@{} LLLLL@{} }
\toprule
Limit cycle Configuration & Average Basin Attraction & SD & Count & Percent \\
\midrule
100001000, 100001110 & 31.50 & 26.57 & 2,658 & 67.29 \\
010111101, 101001100 & 93.75 & 57.41 & 32 & 0.81 \\
010111100, 101001101 & 93.75 & 57.41 & 32 & 0.81 \\
010111011, 101001010 & 93.75 & 57.41 & 32 & 0.81 \\
010111010, 101001011 & 93.75 & 57.41 & 32 & 0.81 \\
010011100, 101101101 & 89.63 & 61.81 & 32 & 0.81 \\
010011010, 101101011 & 89.63 & 61.81 & 32 & 0.81 \\
001101101, 110011100 & 87.78 & 53.75 & 32 & 0.81 \\
001101011, 110011010 & 87.78 & 53.75 & 32 & 0.81 \\
000111101, 111001100 & 81.38 & 44.17 & 32 & 0.81 \\
\bottomrule
\end{tabular*}
\end{table}

\clearpage
\renewcommand{\thepage}{\arabic{page}}


\begin{figure}[ht]
	\centering
	\includegraphics[width=.8\textwidth]{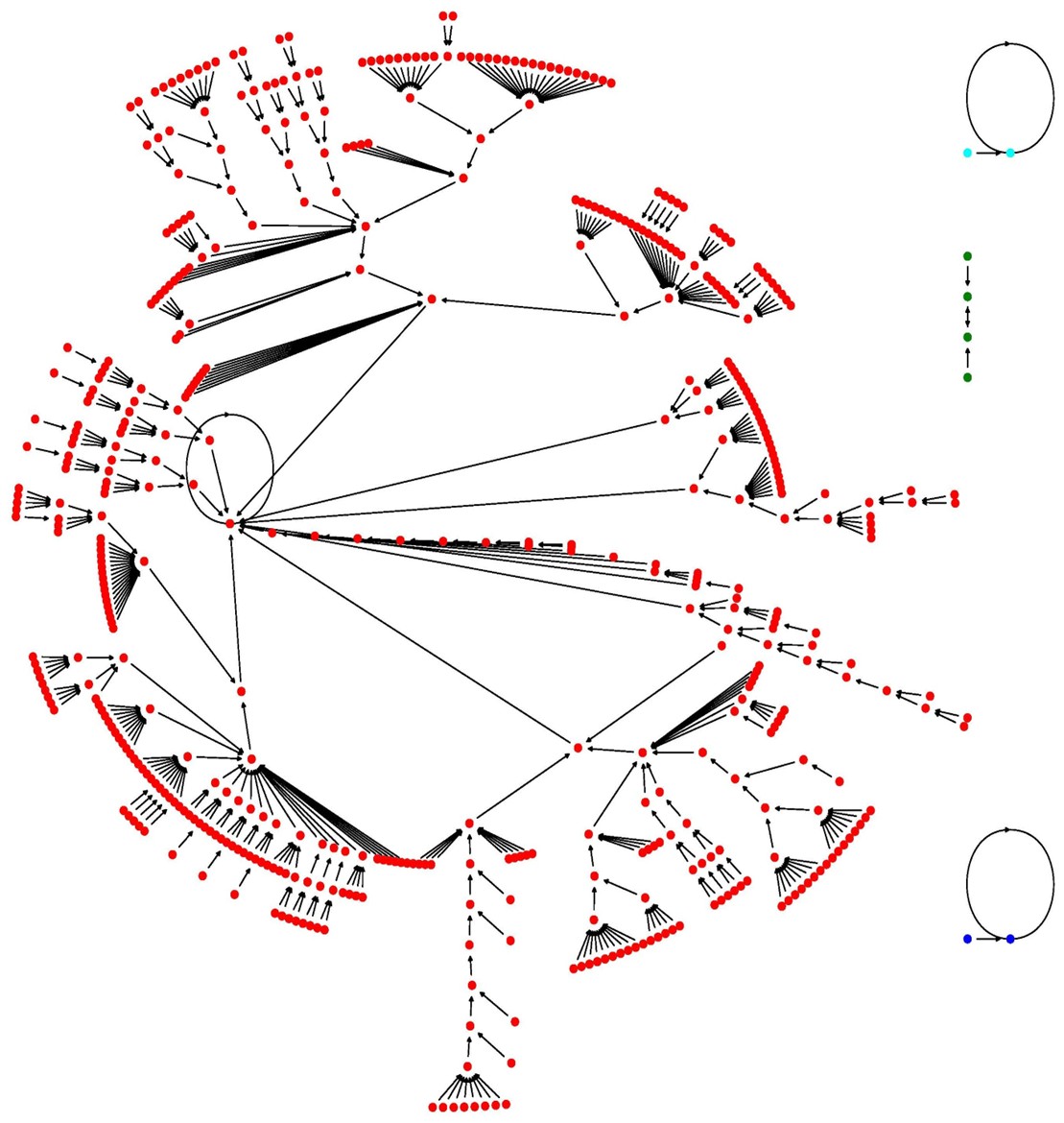}
	\caption{Attraction basin for the synchronous simulation of the 9-node network. This figure presents the attraction basins for the synchronous simulation of the 9-node network. Red points represent the attraction basin for “Drug Resistance”, with a size of 504 configurations. Light blue points represent the attraction basin for “Senescence”, with a size of 2 configurations. Blue points represent the attraction basin for “Apoptosis”, with a size of 2 configurations. Green points represent the attraction basin for the limit cycle, with a size of 4 configurations.}
	\label{FIG:A4}
\end{figure}


\begin{algorithm}[H]
  \caption{FIT\_BOOLEAN\_RULES(Rules)}
  \begin{algorithmic}[1]
    \Require \texttt{Rules}: dictionary mapping each node $\to$ Boolean expression ($\neg,\wedge,\vee$)
    \Ensure \texttt{ValidRules}: dictionary mapping each node $\to$ set of valid candidate rules

    \State \texttt{Nodes} $\gets$ list of all nodes in \texttt{Rules}

    \Statex \(\triangleright\) \textbf{1) Compute original fixed points}
    \State $A_{\text{original}} \gets \emptyset$
    \ForAll{state $s \in \{0,1\}^{|\texttt{Nodes}|}$}
      \If{$\mathrm{EvaluateNetwork}(\texttt{Rules},s) = s$}
        \State $A_{\text{original}} \gets A_{\text{original}} \cup \{s\}$
      \EndIf
    \EndFor

    \Statex \(\triangleright\) \textbf{2) Main loop over each target node}
    \ForAll{target $\in$ \texttt{Nodes}}
      \State \texttt{ValidRules[target]} $\gets \emptyset$
      \State \texttt{Inputs} $\gets \texttt{Nodes} \setminus \{\text{target}\}$

      \Statex \(\triangleright\) \textbf{3) Generate input combinations of size 1–3}
      \For{$r = 1$ to $3$}
        \ForAll{subset $C \subseteq \texttt{Inputs}$ with $|C|=r$}

          \Statex \(\triangleright\) \textbf{4) Generate candidate expressions for $C$}
          \State \texttt{Candidates} $\gets \mathrm{GENERATE\_CANDIDATES}(C)$

          \Statex \(\triangleright\) \textbf{5) Filter by functionality on fixed points}
          \ForAll{expr $\in$ \texttt{Candidates}}
            \If{$\forall\, s\in A_{\text{original}}:\;\mathrm{EVALUATE\_EXPR}(expr,s)=s[\text{target}]$}

              \Statex \(\triangleright\) \textbf{6) Check attractor exactness}
              \State \texttt{Rules'} $\gets$ copy of \texttt{Rules}
              \State \texttt{Rules'}[target] $\gets$ expr
              \State $A_{\text{new}} \gets \emptyset$
              \ForAll{state $t\in\{0,1\}^{|\texttt{Nodes}|}$}
                \If{$\mathrm{EvaluateNetwork}(\texttt{Rules'},t)=t$}
                  \State $A_{\text{new}}\gets A_{\text{new}}\cup\{t\}$
                \EndIf
              \EndFor
              \If{$A_{\text{new}} = A_{\text{original}}$}
                \State \texttt{ValidRules[target]} $\gets \texttt{ValidRules[target]} \cup \{\text{expr}\}$
              \EndIf

            \EndIf
          \EndFor

        \EndFor
      \EndFor

    \EndFor

    \State \Return \texttt{ValidRules}
  \end{algorithmic}
\end{algorithm}

\bigskip

\begin{algorithm}[H]
  \caption{GENERATE\_CANDIDATES($C$)}
  \begin{algorithmic}[1]
    \Require $C$: set of $r$ variables
    \Ensure \texttt{Candidates}: set of expressions

    \State \texttt{Candidates} $\gets \emptyset$
    \If{$|C|=1$}
      \State $v \gets$ the single element of $C$
      \State \texttt{Candidates} $\gets \{\,v,\,\neg v\}$
    \ElsIf{$|C|=2$}
      \State $\{x,y\}\gets C$
      \ForAll{sign\_x $\in\{\text{pos},\text{neg}\}$}
        \ForAll{sign\_y $\in\{\text{pos},\text{neg}\}$}
          \State $\texttt{lit\_x}\gets(\text{sign\_x}=\text{pos})?x:\neg x$
          \State $\texttt{lit\_y}\gets(\text{sign\_y}=\text{pos})?y:\neg y$
          \State \texttt{Candidates} $\gets \texttt{Candidates}\cup\{\texttt{lit\_x}\wedge\texttt{lit\_y},\,\texttt{lit\_x}\vee\texttt{lit\_y}\}$
        \EndFor
      \EndFor
    \ElsIf{$|C|=3$}
      \State $\{X,Y,Z\}\gets C$
      \ForAll{each assignment of negation flags over $\{X,Y,Z\}$}
        \State define \texttt{lit\_X}, \texttt{lit\_Y}, \texttt{lit\_Z}
        \State \texttt{Candidates} $\gets \texttt{Candidates}\cup\{\,\texttt{lit\_X}\wedge\texttt{lit\_Y}\wedge\texttt{lit\_Z},\,\texttt{lit\_X}\vee\texttt{lit\_Y}\vee\texttt{lit\_Z}\}$
        \State \texttt{Candidates} $\gets \texttt{Candidates}\cup\{(\texttt{lit\_X}\wedge\texttt{lit\_Y})\vee\texttt{lit\_Z},\,(\texttt{lit\_X}\vee\texttt{lit\_Y})\wedge\texttt{lit\_Z}\}$
      \EndFor
    \EndIf
    \State \Return \texttt{Candidates}
  \end{algorithmic}
\end{algorithm}

\bigskip
  
\begin{algorithm}[H]
  \caption{EvaluateNetwork(\texttt{Rules}, $s$)}
  \begin{algorithmic}[1]
    \Statex \(\triangleright\)Synchronous update: evaluate each rule with state $s$, return new state $s'$.
  \end{algorithmic}
\end{algorithm}

\bigskip

\begin{algorithm}[H]
  \caption{EVALUATE\_EXPR(expr, $s$)}
  \begin{algorithmic}[1]
    \Statex \(\triangleright\)Substitute literals with values from $s$ and evaluate → Boolean.
  \end{algorithmic}
\end{algorithm}


\begin{table}[width=.9\linewidth,cols=6,pos=ht]
\caption{fitted rules for the 9-node network.}
\label{Atbl10}
\begin{tabular*}{\tblwidth}{@{} LL@{} }
\toprule
Node & Interactions\\
\midrule
miR\_145 & p53 \& !MALAT1 \& !BMI1 \\
Sp1 & (BMI1) | !miR\_145 \\
MALAT1 & Sp1 \\
BMI1 & (!p53\_A \& !p53\_K) | E2F1 \\
KLF4 & !miR\_145 | (E2F1 \& p53) \\
p53 & !KLF4 | !MALAT1 \\
p53\_A & !p53\_K \& p53 \\
p53\_K & !p53\_A \& p53 \\
E2F1 & MALAT1 \\
\bottomrule
\end{tabular*}
\end{table}


\begin{table}[width=.9\linewidth,cols=5,pos=ht]
\caption{Steady state results for the fitted 9-node network simulation.}
\label{Atbl11}
\begin{tabular*}{\tblwidth}{@{} LLLLLL@{} }
\toprule
Components & Steady state 1 & Steady state 2 & Steady state 3\\
\midrule
miR\_145 & 0 & 1 & 1 \\
Sp1 & 1 & 0 & 0 \\
MALAT1 & 1 & 0 & 0 \\
BMI1 & 1 & 0 & 0 \\
KLF4 & 1 & 0 & 0 \\
p53 & 0 & 1 & 1 \\
p53\_A & 0 & 0 & 1 \\
p53\_K & 0 & 1 & 0 \\
E2F1 & 1 & 0 & 0 \\
\bottomrule
Attraction basin & 99.22 & 0.39 & 0.39 \\
\end{tabular*}
\end{table}


\begin{table}[width=.9\linewidth,cols=4,pos=ht]
\caption{Limit cycle analysis for asynchronous update schemes of the fitted 9-node network simulation.}
\label{Atbl12}
\begin{tabular*}{\tblwidth}{@{} LLLL@{} }
\toprule
Steady state Configuration & Average Basin Attraction & SD & Count \\
\midrule
011110001 & 460.87 & 54.98 & 23,107 \\
100001010 & 22.47 & 18.46 & 23,107 \\
100001100 & 22.47 & 18.46 & 23,107 \\
\bottomrule
\end{tabular*}
\end{table}

\clearpage
\renewcommand{\thepage}{\arabic{page}}


\begin{table}[width=.9\linewidth,cols=4,pos=ht]
\caption{Top 10 limit cycle for asynchronous update schemes for the fitted 9-node network simulation.}
\label{Atbl13}
\begin{tabular*}{\tblwidth}{@{} LLLLL@{} }
\toprule
Limit cycle Configuration & Average Basin Attraction & SD & Count & Percent \\
\midrule
010011100, 101101101 & 78.87 & 43.28 & 67 & 3.28 \\
010011010, 101101011 & 78.87 & 43.28 & 67 & 3.28 \\
000111100, 111001101 & 60.42 & 32.22 & 67 & 3.28 \\
000111010, 111001011 & 60.42 & 32.22 & 67 & 3.28 \\
010111010, 101001011 & 88.36 & 52.96 & 67 & 3.28 \\
010111100, 101001101 & 88.36 & 52.96 & 67 & 3.28 \\
000111101, 111001100 & 77.01 & 39.66 & 67 & 3.28 \\
000111011, 111001010 & 77.01 & 39.66 & 67 & 3.28 \\
010111011, 101001010 & 89.07 & 53.82 & 67 & 3.28 \\
010111101, 101001100 & 89.07 & 53.82 & 67 & 3.28 \\
\bottomrule
\end{tabular*}
\end{table}

\end{document}